\documentclass[twocolumn,showpacs,aps,groupedaddress,prd,eqsecnum]{revtex4}

\usepackage{epsfig}
\usepackage{dcolumn}
\usepackage{bm}
\usepackage{latexsym}
\usepackage{amsfonts}
\usepackage{amssymb}

\newcommand{\beq}{\begin{equation}}
\newcommand{\eeq}{\end{equation}}
\newcommand{\beqa}{\begin{eqnarray}}
\newcommand{\eeqa}{\end{eqnarray}}
\newcommand{\bea}{\begin{array}}
\newcommand{\ena}{\end{array}}

\begin{document}
\draft

\title{Considering boundary conditions for black hole entropy in loop quantum gravity}

\author{Takashi Tamaki}
\email{tamaki@gravity.phys.waseda.ac.jp}
\address{Department of Physics, Waseda University, 3-4-1 Okubo,
Shinjuku, Tokyo 169-8555, Japan
}

\date{\today}

\begin{abstract}
We argue for black hole entropy in loop quantum gravity (LQG) by taking into account 
the interpretation that there is no other side of the horizon. This gives new values for 
the Barbero-Immirzi parameter ($\gamma=0.367\cdots$ or $0.323\cdots$) which are fairly larger 
than those considered before ($\gamma=0.261\cdots$ or $0.237\cdots$). We also discuss 
its consequences for future experiments. 
\end{abstract}

\pacs{04.70.Bw, 04.30.Db, 04.70.Dy} \maketitle

\section{Introduction}
Black hole thermodynamics is one of the most exciting arenas for those investigating 
quantum gravity. In particular, discovery of the microscopic origin of black hole entropy 
in string theory has attracted much attention~\cite{Strominger}. However, its understanding is 
basically restricted by the perturbative formulation of string theory. 
As a result, the origin of black hole entropy has been revealed only for BPS black holes. 

In this respect, the explanation of the origin of black hole entropy 
in loop quantum gravity (LQG), which has background-independent formulation, 
is stimulating~\cite{Corichi}. This explanation was attempted based on 
spin network formalism~\cite{Smolin}. In LQG, it has also been 
reported that we can avoid initial or final singularities in the 
universe~\cite{Bojo} and the singularity in black holes~\cite{Bojo2}. One of the drawbacks in explaining 
black hole entropy is that 
there is a free parameter called the Barbero-Immirzi parameter $\gamma$, which is related to an ambiguity 
in the choice of canonically conjugate variables~\cite{Barbero-Immirzi}. This is why we can adjust 
the parameter to obtain the relation that black hole entropy $S$ is equal to a quarter 
of the horizon area $A$, i.e., $S=A/4$. Thus, it is important to obtain a consensus about 
how to count the number of degrees of freedom. 

At present, various possibilities have been argued with respect to this 
freedom~\cite{Domagala,Meissner,Alekseev,Khriplovich,Mitra,Corichi2,Tamaki2}. 
However, there is one possibility that has not been considered so far. 
For example, general expressions for the area spectrum $A$ are~\cite{Rovelli,Ash1} 
\beqa
A=4\pi \gamma \sum \sqrt{2j_{i}^{u}(j_{i}^{u}+1)+2j_{i}^{d}(j_{i}^{d}+1)
-j_{i}^{t}(j_{i}^{t}+1)}\ . 
\nonumber
\eeqa
The sum is the addition of all intersections between a surface and edges. 
Here, the indices $u$, $d$, and $t$ mean edges above, below, and tangential to the 
surface, respectively (We can determine which side is above or below arbitrarily). 
So far the number of states of black holes has been discussed based on the simplified 
area spectrum ($j_{i}^{u}=j_{i}^{d}:=j_{i}$ and $j_{i}^{t}=0$). 
Another interesting possibility, however, is to take  
\beqa
j^{d}_{i}=0\ ,
\label{one-way}
\eeqa
to reflect the absence of the other side of the horizon~\cite{Thiemann}. 
It is assumed that a horizon (emerging classical null surface) is a 2 dimensional surface 
that is tangential to the selected edges. In the absence of a direct derivation of 
the classical regime in LQG it is a legitimate speculation.
In this case, the area spectrum becomes 
\beqa
A=4\pi \gamma \sum \sqrt{j_{i}(j_{i}+1)}\ ,
\label{AreaLQG2}
\eeqa
since we have $j_{i}^{u}=j_{i}^{t}:=j_{i}$. 
This is important since it would affect the number of states (and the 
resulting Barbero-Immirzi parameter), which we discuss in this paper. 

Our motivation using (\ref{one-way}) is based on the calculation in the 
entanglement entropy approach~\cite{Nielsen,Bombelli}. I.e., if we express the wave 
function $\Psi$ by a product as 
\beqa
\Psi :=\Psi_{\rm O}\Psi_{\rm H}\Psi_{\rm I}, 
\label{density}
\eeqa
where the indices O, H, and I mean outside, at, and 
inside the horizon, respectively, then, we can construct a density matrix. 
Since the inside of a black hole is inaccessible for an asymptotic observer, 
one traces over $\Psi_{\rm I}$ to calculate entropy~\cite{foot2}. This means that 
we can determine black hole entropy independent of the inside of the horizon. 
This view point sometimes appears as a holographic principle~\cite{'tHooft} and 
is expressed in AdS/CFT corespondence~\cite{AdS}. 
Stimulated by this insight, we proceed by assuming that a black hole area is independent of the spin 
inside the horizon. (Note also that the tangential edges stay tangential for an 
asymptotic observer. This is important for (\ref{AreaLQG2}) and is self-consistent with the 
assumptions mentioned above.) 
Of course, this implicitly assumes a relation between the 
area and the entropy. Thus, it is crucial to consider a method to justify its consequences. 
This is discussed in the final section. 

This paper is organized as follows. In Sec.~II, we summarize and reconsider the framework 
\cite{Corichi} (which we call the ABCK framework.) that is necessary in counting the number 
of states of black holes. In Sec.~III, we determine the number of states. 
In Sec.~IV, we summarize our results and discuss their meaning. 

\section{Consideration of the ABCK framework}

Here, we briefly introduce and consider the ABCK framework in Ref.~\cite{Corichi}. 
One usually considers the event horizon, which is determined after the complete evolution 
of space-time, when one describes black hole thermodynamics. Thus, it would be too 
restrictive to establish a thermodynamical situation in which the system is isolated. 
To explore this idea appropriately, the isolated horizon (IH) is defined in the ABCK framework.
The main difficulty in defining 
IH is to establish the surface gravity or black hole thermodynamics when, in general, 
there is no global Killing field. For details, see~\cite{isolated}. 
Because of the requirement at IH, we can reduce the SU(2) connection to the U(1) connection.  
Using the curvature $F_{ab}$ of the U(1) connection, we can express the boundary condition 
between IH and the bulk as 
\begin{equation}
F_{ab}=-\frac{2\pi \gamma}{A}\underline{\Sigma}_{ab}^{i}r_{i}\ , \label{IH-classical}
\end{equation}
where $A$ is the area of IH. $\Sigma_{ab}^{i}$ is related to a triad density $E_{i}^{a}$ as 
\begin{equation}
E_{i}^{a}=\gamma\eta^{abc}\Sigma_{bci}\ , \label{sigma}
\end{equation}
where $\eta^{abc}$ is the Levi-Civita 3-form density. $\underline{\Sigma}_{ab}^{i}$ is its 
pull back to IH and $r_{i}$ is unit normal. (\ref{IH-classical}) plays an important role 
in determining the condition (iv) below. 

Usually, we consider the Hilbert space using the spin network in LQG. 
When there is IH, we decompose the Hilbert space as the tensor product of that in  
IH $H_{\rm IH}$ and that in the bulk $H_{\Sigma}$, i.e., $H_{\rm IH}\otimes H_{\Sigma}$. 

First, we consider $H_{\Sigma}$. Using edges having spin $(j_{1},j_{2},\cdots ,j_{n})$ 
which pierce IH, we can write $H_{\Sigma}$ as the orthogonal sum 
\begin{equation}
H_{\Sigma}=\bigoplus_{j_{i},m_{i}}H_{\Sigma}^{j_{i},m_{i}}\ , \label{sum}
\end{equation}
where $m_{i}$ takes the value $-j_{i}$, $-j_{i}+1$, $\cdots$, $j_{i}$. 
This is related to the flux operator eigenvalue $e_{s'}^{m_{i}}$ 
that is normal to IH ($s'$ is the part of IH that has 
only one intersection between the edge with spin $j_{i}$.)
\begin{equation}
e_{s'}^{m_{i}}=8\pi\gamma m_{i}\ . \label{flux}
\end{equation}
Then, when we argue the possibility (\ref{one-way}), it is natural to 
restrict as $m_{i}> 0$ since $m_{i}< 0$ corresponds to the flux operator eigenvalue 
inside the horizon. Basically, constraints in the bulk do not affect the number  
counting, as shown in ~\cite{Corichi}. 
The area operator eigenvalue $A_{j}$ should satisfy~\cite{foot1} 
\beqa
{\rm (i)}\ \ A_{j}=4\pi \gamma \displaystyle  \sum_{i} \sqrt{j_i(j_i+1)}\leq A\ . 
\label{condition1}
\eeqa

Next, we consider $H_{\rm IH}$. We have, in general, difficulty in constructing $H_{\rm IH}$. 
However, if we fix the horizon area $A$ as 
\begin{equation} 
A=4\pi \gamma k\ \ , \label{area} 
\end{equation} 
where $k$ is a natural number, which is called the level of the Chern-Simons theory, 
we can construct $H_{\rm IH}$ using a function which is invariant 
under the diffeomorphism and the $Z_{\rm k}$ gauge transformation, i.e., a 
``quantized" U(1) gauge transformation. In addition to this condition, 
it is required that 
\begin{equation} 
{\rm (ii)\ we\ should\ fix\ an\ ordering}\ (j_{1},j_{2},\cdots ,j_{n}). \label{ordering} 
\end{equation} 
At IH, we do not consider the scalar constraint, since the lapse function disappears. 
As a result, $H_{\rm IH}$ is written as an orthogonal sum similar to (\ref{sum}) 
by eigenstates $\Psi_{b}$ of the holonomy operator $\hat{h}_{i}$, i.e., 
\beqa
\hat{h}_{i}\Psi_{b}=e^{\frac{2\pi ib_{i}}{k}}\Psi_{b}\ . 
\label{holonomy}
\eeqa
From the quantum Gauss-Bonnet theorem that guarantees that IH is $S^{2}$, 
we require 
\beqa
{\rm (iii)}\ \ \sum_{i=1}^{n}b_{i}=0\ \ \ {\rm mod}\ k\ . 
\label{GB}
\eeqa

Finally, we should consider the quantization of the boundary condition between IH and the bulk 
(\ref{IH-classical}). Since only the exponential version $exp(i\hat{F})$ is welldefined on 
$H_{\rm IH}$, we consider 
\begin{equation}
(exp(i\hat{F})\otimes 1)\Psi =(1\otimes \exp(-i\frac{2\pi \gamma}{A}\underline{\Sigma}\cdot r))
\Psi \ , \label{IH-quantum}
\end{equation}
where $\Psi$ expresses the state in $H_{\rm IH}\otimes H_{\Sigma}$. 
From this, we have 
\beqa
{\rm (iv)}\ \ b_{i}=-2m_{i}\ \ \ {\rm mod}\ k\ . 
\label{boundary}
\eeqa
All we need to consider in number counting is (i)-(iv).

\section{number counting}

Here, we consider number counting based on the ABCK framework. 
If we use (iii) and (iv), we obtain 
\beqa
{\rm (iii)}'\ \ \sum_{i=1}^{n}m_{i}=n'\frac{k}{2}\ . 
\label{GB2}
\eeqa
In \cite{Meissner}, it was shown that this condition is irrelevant in number 
counting. Thus, we perform number counting concentrating only on (i) and (ii) below. 
There are two opinions in number counting. 
The one adopted in the original paper \cite{Corichi,Domagala,Meissner} 
counts the {\it surface} freedom 
$(b_{1},b_{2},\cdots ,b_{n})$. The second 
counts the freedom for both $j$ and $m$ \cite{Khriplovich,Mitra,Tamaki2}. 
Here, for simplicity, we base our argument mainly on the second possibility. 
We can perform this calculation in a manner quite analogous to \cite{Tamaki2}. 
The only point in which care must be taken is that we should include 
the condition $m_{i}> 0$ (or equivalently $b_{i}< 0$). Thus, 
each $j_{i}$ has freedom $j_{i}$ for the $j_{i}$ integer and the $j_{i}+1/2$ way 
for the $j_{i}$ half-integer. They are 
summarized as $[\frac{2j+1}{2}]$, where $[\cdots ]$ is the integer part. 
We define $N(a)$ ($a:=\frac{A}{8\pi\gamma}$), which is the number of states accounting 
for the entropy as 
\beqa
\hspace{-5mm}N(a):=\left\{(j_{1},\cdots, j_{n})|0\neq j_{i}\in \frac{N}{2}, \right. 
\nonumber \\
\hspace{-5mm}\left.\displaystyle  \sum_{i} \frac{\sqrt{j_i (j_i +1)}}{2}\leq 
\frac{k}{2}=a
\right\}\ . 
\label{number2}
\eeqa
Then, we obtain the recursion relation 
\beqa
\hspace{-10mm}&&N (a)=\left\{N\left(a-\frac{\sqrt{3}}{4}\right)-1\right\}+
\left\{N\left(a-\frac{\sqrt{2}}{2}\right)-1\right\} 
\nonumber  \\
\hspace{-10mm}&&+\cdots+\left[\frac{2j_{i}+1}{2}\right]\left\{N\left(a-\frac{\sqrt{j_i (j_i +1)}}{2}
\right)-1\right\}+\cdots 
\nonumber  \\
&&+[\sqrt{16a^{2}+1}+1]\ . 
\label{recursion2}
\eeqa
The factor $[\frac{2j_{i}+1}{2}]$ in this formula comes directly from the freedom of $j_{i}$. 
If we use the relation
\beqa
N(a)=Ce^{ \frac{A\gamma_{M}}{4\gamma} }\ , \label{number-behavior}
\eeqa
where $C$ is a constant, we obtain 
\beqa
1=\sum_{j_{i}=Z/2}[\frac{2j_{i}+1}{2}]\exp (-\pi\gamma_{M} \sqrt{j_{i}(j_{i}+1)})\ 
\label{Barbero-Immirzi4}
\eeqa
by plugging (\ref{number-behavior}) into (\ref{recursion2}) and 
taking the limit $A\to\infty$. Then if we require $S=A/4$, we have 
$\gamma =\gamma_{M}=0.367\cdots$. Quite analogously, if we count the surface freedom, we obtain
\beqa
1=\sum_{j=Z/2}\exp (-\pi\gamma_{M} \sqrt{j(j+1)})\ ,
\label{Barbero-Immirzi5}
\eeqa
where we have $\gamma_{M}=0.323\cdots$. Importantly, these are fairly larger values 
compared with those of~\cite{Domagala,Meissner,Alekseev,Khriplovich,Mitra,Corichi2,Tamaki2}.

\section{Conclusion and discussion}

In this paper, we have considered condition (\ref{one-way}) to derive 
the number of states of black holes in the ABCK framework. As a result, we obtained 
two values of the Barbero-Immirzi parameter according to how the freedom is counted. 
Importantly, these are fairly larger values compared with previous 
results~\cite{Domagala,Meissner,Alekseev,Khriplovich,Mitra,Corichi2,Tamaki2}. 

What are the consequences of this result? 
The first one is related to cosmology~\cite{LQC}. If we assume isotropic and 
homogeneous universe, we can write the effective inverse cube of scale factor as 
\beqa
(a^{-3})_{\rm eff}=a^{-3}f_{l}(q)^{3/(2-2l)}\ ,
\nonumber
\eeqa
where
\beqa
&&f_{l}(q)=3q^{1-l}/(2l)\left\{
(l+2)^{-1}[(q+1)^{l+2}-|q-1|^{l+2}]-\right.  \nonumber  \\
&&\left.(l+1)^{-1}q[(q+1)^{l+1}-{\rm sgn} (q-1)|q-1|^{l+1}]
\right\}\ ,
\nonumber
\eeqa
and where $q=a^{2}/a_{\ast}^{2}$ ($a_{\ast}^{2}=\gamma l_{pl}^{2}j/3$). $j$ and $l$ are 
the spin and the ambiguity in quantization. We can argue inflationary scenario 
using the following scalar field effective action
\beqa
H_{M}=\frac{(a^{-3})_{{\rm eff}}}{2}p_{\phi}^{2}+a^{3}V(\phi )\ .\nonumber 
\eeqa
Thus, the result certainly depends on the value of the Barbero-Immirzi parameter. 
This would affect the thermal fluctuation in the universe, and might be seen by, e.g., PLANCK. 

As other possibilities, we can also discuss particle velocity and its dependence 
on the Barbero-Immirzi parameter~\cite{LIV}, or discuss the evaporation process of black holes~\cite{eva}. 
Thus, the value of the Barbero-Immirzi parameter should also be certified in future experiments. 

\acknowledgements
We would like to thank Hidefumi Nomura, Lee Smolin, and Olaf Dreyer for useful discussion. 
The numerical calculations were carried out on the Altix3700 BX2 at YITP, Kyoto University. 
This work was partially supported by the 21st Century COE Program (Holistic Research and 
Education Center for Physics Self-Organization Systems) at Waseda University.

%


\begin{references}
\bibitem{foot2}
See, \cite{Livine} which argued for yet another possibility of the degrees-of-freedom counting 
using entanglement considerations in loop quantum gravity.
\bibitem{foot1}
Though we consider $A$ not an interval $[A-\delta A, A+\delta A]$, 
this does not affect the final results. Because of the expression 
(\protect\ref{number-behavior}), $S:=\ln W:=\ln (\frac{dN}{dA}\delta A)$ 
is equal to $\ln N$ for $A\to\infty$.
\end{references}
\end{document}